\documentclass[]{interact}
\usepackage{mathtools}
\usepackage{algorithm}
\usepackage{placeins}
\usepackage{color,soul}

\usepackage{epstopdf}
\usepackage[caption=false]{subfig}
\def\bea{\begin{eqnarray}}
\def\eea{\end{eqnarray}}
\def\be{\begin{equation}}
\def\ee{\end{equation}}

\usepackage[natbibapa,nodoi]{apacite}
\theoremstyle{plain}

\theoremstyle{definition}

\theoremstyle{remark}

\begin{document}


\title{Control-Oriented, Data-Driven Models of Thermal Dynamics}

\author{
\name{Ljuboslav Boskic\textsuperscript{a}\thanks{Email: lboskic@ucsb.edu} and Igor Mezi\'{c}\textsuperscript{a}}
\affil{\textsuperscript{a}Department of Mechanical Engineering, University of California Santa Barbara}
}
\maketitle

\begin{abstract}
We investigate data-driven, simple-to-implement residential environmental models that can serve as the basis for energy saving algorithms in both retrofits and new designs of residential buildings.  Despite the nonlinearity of the underlying dynamics, using Koopman operator theory framework in this study we show that a linear second order model embedding, that captures the physics that occur inside a single or multi zone space does well when compared with data simulated using EnergyPlus.  This class of models has low complexity. We show that their parameters have physical significance for the large-scale dynamics of a building and are correlated to concepts such as the thermal mass. We investigate consequences of changing the thermal mass on the energy behavior of a building system and provide best practice design suggestions.  
\end{abstract}

\begin{keywords}
Energy Efficiency, Residential Buildings, Reduced Order Model
\end{keywords}

\section{Introduction}
\indent The ``House as a System'' approach is gaining traction as a protocol to gain deep energy efficiency in residential buildings (\cite{hoickaetal:2018}). However, the current approach is focused on scheduling the order of retrofits (insulation first, replacement of furnace second, etc.) and thus high capital expenditure actions. In commercial buildings, the cost of such retrofits has led to development of strategies for optimizing operations of existing systems, focusing first on fault detection and returning the building operation to a ``healthy" state (\cite{Littooyetal:2016}). Beyond the fault detection methodologies, model-based approaches lead to optimization of existing systems and potential of deep energy savings for new commercial builds (\cite{georgescuandmezic:2015}), and even US Army facilities (\cite{Mezic:2017}). However, these gains are not currently utilized in the context of residential buildings. 

\indent Therefore, residential buildings have recently gained more attention within the topic energy efficiency. There are about 136.5 million residential buildings in the United States (\cite{USHomes}), creating a large opportunity for energy savings via retrofits and new designs, to create more efficient homes. Through addition of sensing, communication and actuation of components, devices are made "smart", such that they communicate wirelessly with each other and transmit data to help reduce use during peak demand periods. With energy monitoring and cost savings, smart home technologies have potential to deliver benefits such as convenience, control, security and monitoring, environmental protection, and simply enjoyment from engaging with the technology itself (\cite{ford2017}). In order for retrofits and newly designed systems to work properly, smart technology must be introduced and implemented. Smart technology incorporates sensors, actuators and algorithms. Here we present a reduced order model (ROM) for indoor temperature of a single zone, with the goal of improving energy efficiency for residential buildings that can serve as a basis for all energy saving algorithms which requires no cloud computing.  \

In this work we present a modeling approach that takes a global, physical point of view. Namely, based on model order reduction ideas emerging from Koopman operator theory (\cite{Mezic:2005}), we introduce a class of linear second order thermal models. Work using Koopman operator theory has been done and proved as a valuable incite in deriving our mode (\cite{Boskic2020,Masaki_2019}), We show that the model coefficients reflect well-known physical properties such as the thermal mass, thermal damping, conduction and radiation. Despite the nonlinearity of the underlying dynamics, in this study we show that a linear second order model embedding, that captures the physics that occur inside a single or multi zone space does well when compared with data simulated using EnergyPlus. This class of models has low complexity. We show that their parameters have physical significance for the large-scale dynamics of a building and are correlated to concepts such as the thermal mass. We investigate consequences of changing the thermal mass on the energy behavior of a building system and provide best practice design suggestions.  
 
The paper is organized as follows: In section \ref{sec:E+} we introduce the Energy Plus model that we use to validate the reduced order model. In section \ref{sec:Koop} we derive the general form of a reduced order model using Koopman operator theory and discuss the physical meaning of its parameters. In section \ref{sec:single} we test the performance of the reduced order model against the Energy Plus simulation of a single-zone building. In section \ref{sec:multi} we do the same for a multi-zone building.

\FloatBarrier
\begin{section}{Model Description}
 \label{sec:E+}
\indent The residential building model used in the analysis was constructed in Sketchup (Computer-aided design (CAD) software) \cite{Sketchup} and then applied using OpenStudio \cite{Openstudio} and EnergyPlus software  to run a year long simulation. The location used in this study is Santa Barbara, California. The outdoor temperature for Santa Barbara is obtained from the Department of Energy EnergyPlus website for the year  2009. The model zone of a building, as seen in figure \ref{House} has dimensions of $7.72m\times7.72m\times3.046m$ with an approximate volume of $181.5m^3$. The building has 3 windows and one door. All the material used is based on ASHRAE 189.1 standard corresponding to the location of the test area. \
\begin{figure} [h!]
\centering
\includegraphics[scale=0.25]{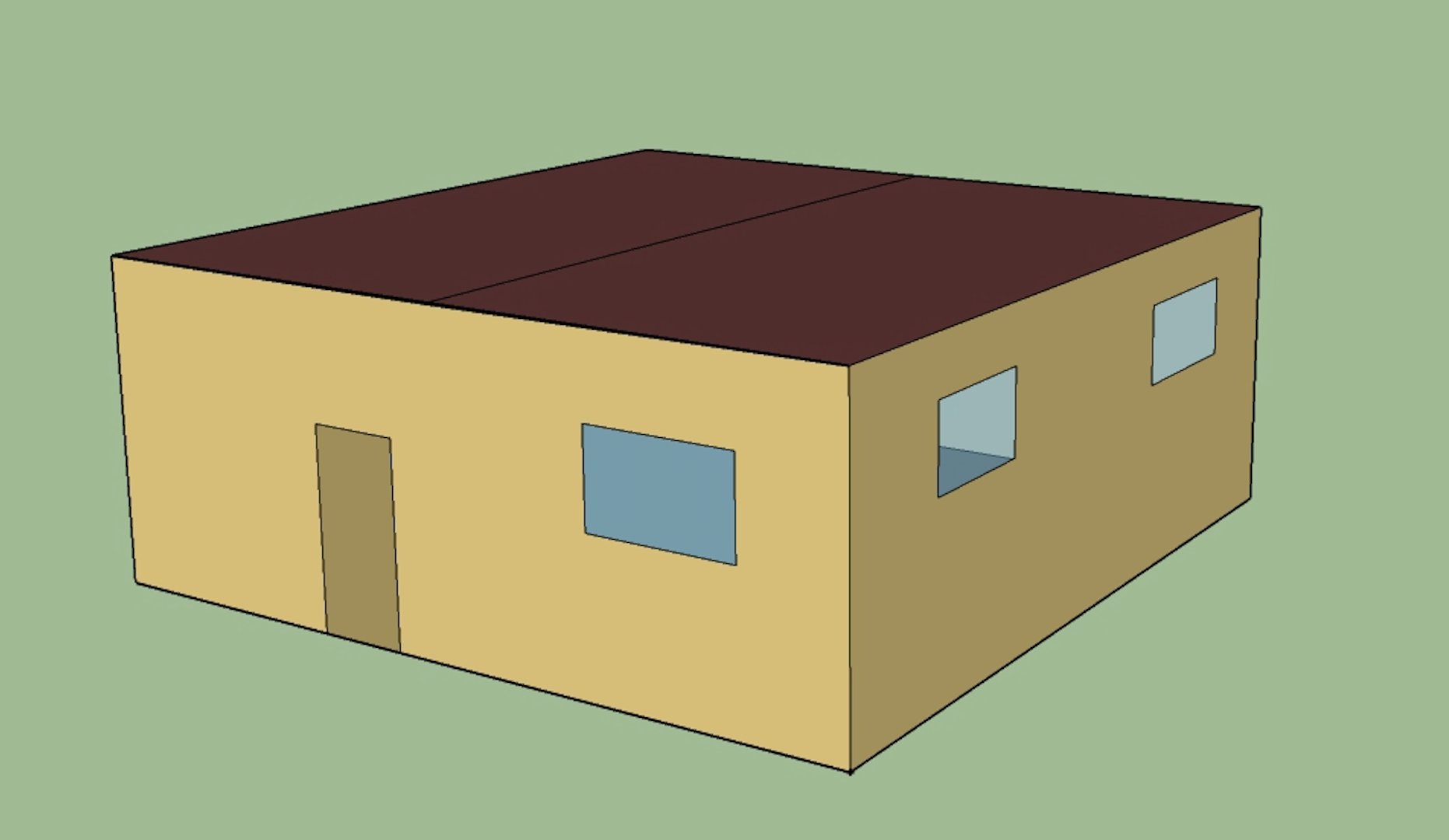}
\caption{Sketchup constructed model of single thermal zone house.}
\label{House}
\end{figure}

\indent In OpenStudio,  a wide-variety of conditions such as setpoints, occupant schedules, HVAC equipment, loads, and more (see figure \ref{workflowOpenstudios}) can be specified. We first developed the nominal, no-actuation, no-load  model, that enables us to parametrize important physical concepts such as the thermal mass by turning off all thermal loads. The model outputs were compared with a reduced order model the development of which we describe next.

\begin{figure} [h!]
\centering
\includegraphics[scale=0.25]{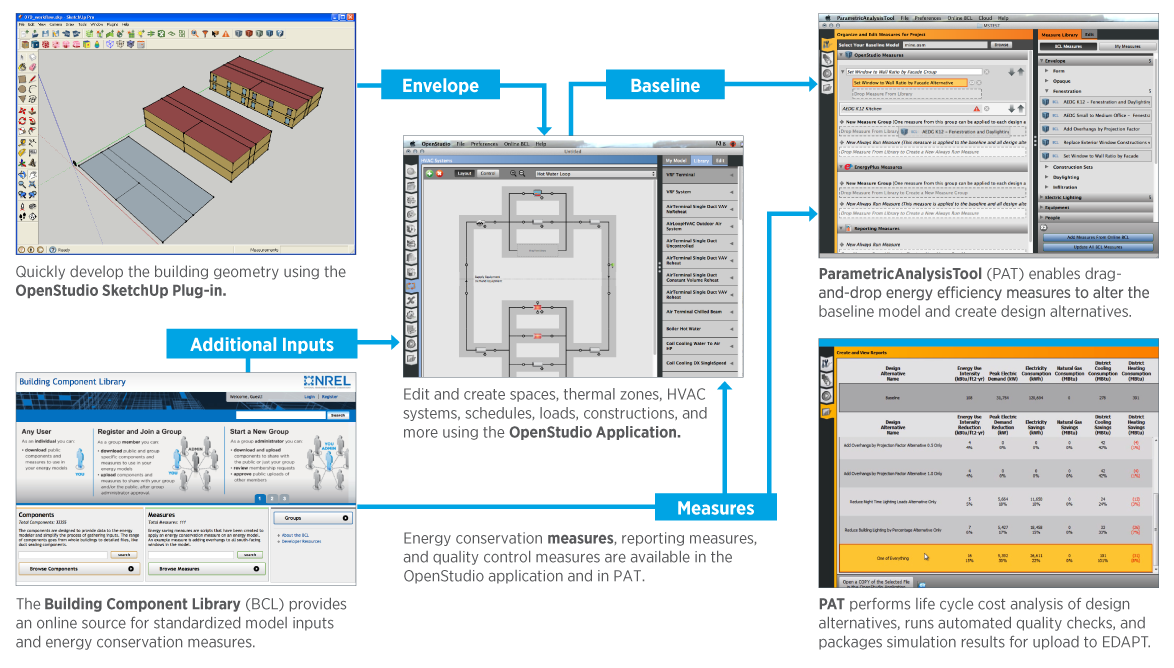}
\caption{Openstudio work flow }
\label{workflowOpenstudios}
\end{figure}

\end{section}
\FloatBarrier
\begin{section}{Reduced Order Model}\
\indent 
\label{sec:Koop}
There are a number current attempts to use modeling approaches for control in commercial buildings (\cite{may2011model,avci2013model,hazyuk2012optimal}). These are typically based on  thermal models that are attempting to capture the details of all the thermal interactions in buildings (\cite{hazyuk2012optimal}). This leads to high complexity of the models, making them less likely for implementation without cloud computation, as well as lack of insight into the global properties of the thermal dynamics. Moreover, to transfer such technology to residential building, the model underlying control has to be simple, computable ``at the edge" instead on the cloud and amenable to exploit innovative control actuation such as active thermal mass control.

The model of building physics we develop here is  simple,  yet it captures the relevant large-scale physical effects. Our methodology is inspired by data-driven approach to control utilizing Koopman operator methods. Starting from papers (\cite{mezic2004comparison,Mezic:2005}) these methods gained widespread adoption in fields as diverse as fluid mechanics (\cite{mezic2013analysis}), power grid (\cite{susuki2016applied}) and control theory (\cite{mauroy2019koopman}). The theory utilizes Koopman operator eigenfunctions to develop linear reduced order models of dynamical systems. Namely, an eigenfunction $z$ of the Koopman operator satisfies

\be
\dot {z}=\lambda z=(\sigma+i\omega)  {z},
\ee
where $\lambda$ is the associated eigenvalue.
Then
\be
\ddot {z}=(\sigma^2-\omega^2){z}+ i2\sigma\omega  {z},
\ee

Now,
\bea
\ddot {z}&=&(\sigma^2-\omega^2){z}+ i2\sigma\omega  {z}\nonumber \\
&=&(\sigma^2-\omega^2){z}+ 2\sigma(\dot {z}-\sigma z)\nonumber \\
&=&-(\sigma^2+\omega^2)z+ 2\sigma\dot {z}
\eea
In viscously damped vibrations such equations are used, and special values of $\sigma$ and $\omega$ are used to obtain 
the real solution, that follows from a second order equation that involves a real, velocity-dependent damping force.
Namely, if we require
\[
\omega_n^2=\sigma^2+\omega^2,
\]
and
\[
\sigma=-\xi\omega_n,
\]
we get
\[
\omega_n^2=\sigma^2+\omega^2=\xi^2\omega_n^2+\omega^2,
\]
which implies
\[
\omega^2=(1-\xi^2)\omega_n^2,
\]
the classical viscous damping result. We can also observe $\sigma=-c/2m$, and $\xi=-c/2k$ where $c$ is the damping coefficient, $k$ is the stiffness and $m$ the mass of the vibrating system.

Motivated by the above discussion, we make an assumption that the temperature dynamics of a building can be represented by its first Koopman mode, thus 
obtaining  a second order linear differential equation with constant coefficients as our model for temperature inside a particular space/thermal zone. Labeling the state $x$ as the temperature, the equation reads, 
\begin{align*}
c_1& \ddot{x}+c_2\dot{x}+c_3x =u+c_4, 
\end{align*}
where $u$ is external input, or 
\begin{equation} \label{EQ1}
\ddot{x} =-\frac{c_3}{c_1}\dot{x}-\frac{c_2}{c_1}x+\frac{1}{c_1}u+\frac{c_4}{c_1},
\end{equation}\
It is intuitive from the discussion above that $c_1$ should represent the ``mass" parameter, in this thermal model  being the thermal mass.
We rewrite the equation in a state space representation:  \

\begin{equation} \label{EQ2}
\dot{x}= \begin{bmatrix} 0 & 1 \\ \frac{-c_3}{c_1} & \frac{-c_2}{c_1} \end{bmatrix}x + 
\begin{bmatrix}
0 \\ \frac{1}{c_1}
\end{bmatrix}u + 
\begin{bmatrix}
1 \\ 0
\end{bmatrix}
\frac{c_4}{c_1}
\end{equation}\
 
 We now discuss what every term in equation (\ref{EQ2}) represents in terms of thermal physics of the space. The thermal mass influences the $c1$ term strongly. Thermal mass is a very important aspect in buildings due to it being the main source of absorption of outside and inside thermal and passive control of the living space inside. The thermal mass is influenced by the physical structure of the walls of the building, because of the varying ability of the material to absorb and store heat energy. For example, a lot of thermal energy is needed to change the heat inside a building that has been constructed out of brick, due to the fact that the density of the material is high. In fact, any material that has greater thermal mass can store more heat and therefore it will take longer to release the thermal energy after the heat source or the sun is gone. \

 Thermal insulation affects the "damping term" $c2$.   It is used to reduce heat loss or gain by providing a barrier between areas that are significantly different in temperature. Insulation is commonly added between the outside walls and inside walls of the house, this is what provides that barrier of protection from the sun. Insulation and thermal mass both slow down the movement of heat between exterior and interior space. Insulation is used when a desired temperature differential is wanted between the indoor and outdoor space. Thermal mass is inertial, as it involves a substance that will slowly take on heat and then slowly release it over time (\cite{InsulationvsThermal}). 

 Heat conduction affects the $c3$ term. Thermal conduction happens when internal energy or heat is transferred by collision of particles and movement of electrons. Material within the walls have different heat conduction properties. The coefficient $c_3$ affects is in a sense a ``global" heat conduction coefficient. Changing the materials in the walls affects  both the thermal mass, and the thermal conduction term. 

 Thermal radiation, the heat transferred by electromagnetic waves such as the visible light or transfer of heat within or through two bodies  affects the term $c4$ in our equation. It was shown in (\cite{radiation}) that radiation heat transfer results in an increase in the heat transfer rate reflecting significant radiation effects that contribute to less thermal resistance. 

 We note that the coefficients above are also affected by factors such as   the orientation of the building. Thus, various physical and design considerations affect the coefficients in a heterogeneous way.
{\it Roughly, thermal mass affects $c1$, insulation affects $c2$, heat conduction coefficients $c3$, and thermal radiation $c4$.} 
The above  Reduced Order Model (ROM) (\ref{EQ1})  reduces computational complexity, from a computationally expensive EnergyPlus simulation to the simple model that can be implemented using embedded controllers and has all the essential physics encoded in its coefficients. Having ROM's it is also easier to understand the nature of systems due to its simplicity. 
\end{section}

\FloatBarrier

\begin{section}{Results for Single Zone Model}
\label{sec:single}
In this section we report results obtained using a single-zone model.
 In order to illustrate some of the complexity of temperature changes, in figures \ref{inout1} and   \ref{inout2} we provide the indoor and outdoor temperature plots for one case of 286 operational hours.   There is a noticeable shift (delay) between the peaks of temperature between the  outdoor and indoor temperatures. Control of the shift  can be done e.g. by  the old fashioned manual implementation of opening the window before the sun is out and closing it afterwords in order to cool the home. However, the complexity of the delay timing indicates an automatic controller would be better in determining the exact times for such action, especially if the actuation is done using non-standard means such as active thermal mass. 
\begin{figure}[ht!]
    \centering
        \includegraphics[scale=0.5]{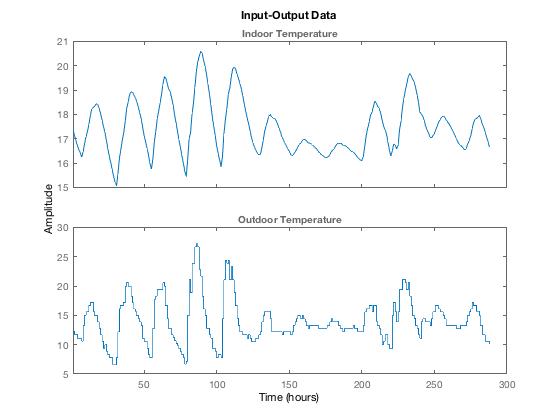}
 		\caption{Hourly indoor-outdoor temperature plots during 2/26 to 3/09}
        \label{inout1}
\end{figure}

\begin{figure}[ht!]
\centering
        \includegraphics[scale=0.5]{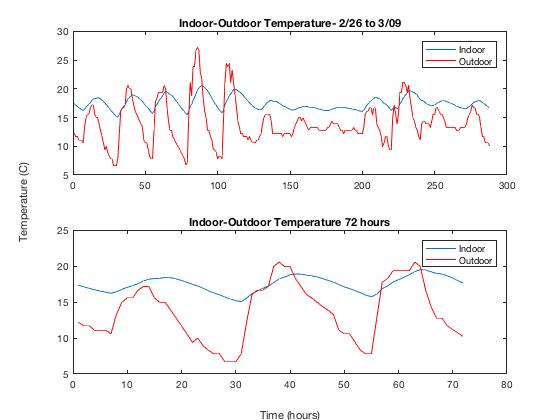}
 		\caption{Indoor-outdoor temperature plot of 72 hour close up.}
        \label{inout2}
\end{figure}
\indent We implement  system identification technique to get the optimal coefficients  described in the previous section. The modeled indoor temperature compared to the "actual" indoor temperature from that particular zone from the EnergyPlus simulation is shown in figure  \ref{Model286}. The percentage error found between the actual indoor temperature from simulation to our model is $6.3512\%$ showing very good performance of the reduced order model with optimized coefficients. The error was calculated using root mean square error. 

\begin{figure}[ht!]
\centering
        \includegraphics[scale=0.6]{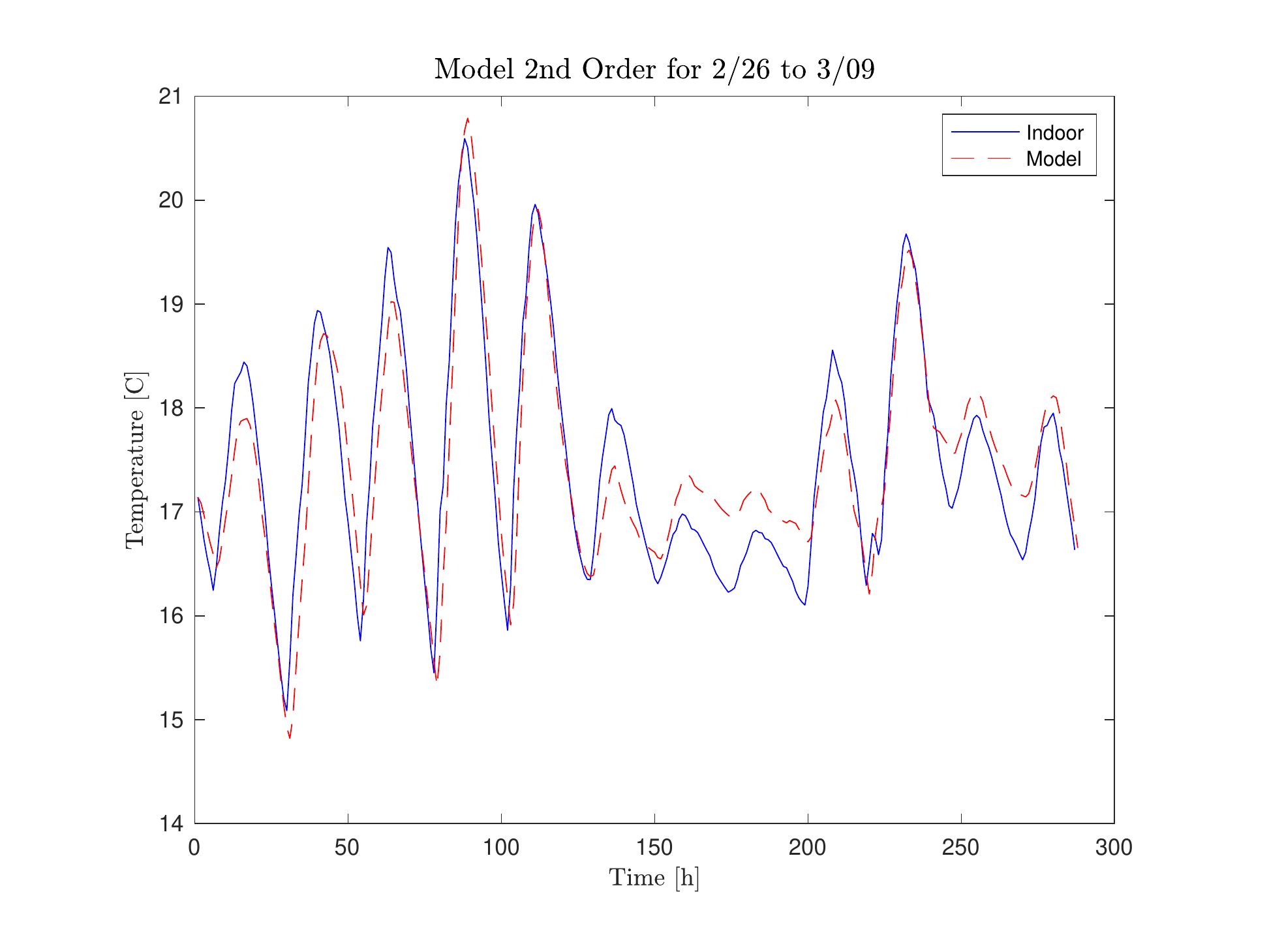}
 		\caption{ROM for indoor air temperature from 2/26 to 3/09}
        \label{Model286}
\end{figure}
\end{section}

\FloatBarrier
\begin{section}{Results for Multi-Zone Model}
\label{sec:multi}
\indent In the previous section we tested the reduced order modeling approach using a  single zone building.In this section we analyze performance for a multi-zone building
whose features are shown in figures \ref{MultizoneModel}, \ref{SpaceType} and \ref{ThermalZone}. In EnergyPlus  we had four separate thermal zones and then we added a single thermal zone for the whole house that in a sense computes a weighted average of the four thermal zone spaces. Note that  zoning based on thermal properties of neighboring spaces based on Koopman operator methods was done in (\cite{georgescu2012creating}).

\begin{figure}[ht!]
\centering
        \includegraphics[scale=0.15]{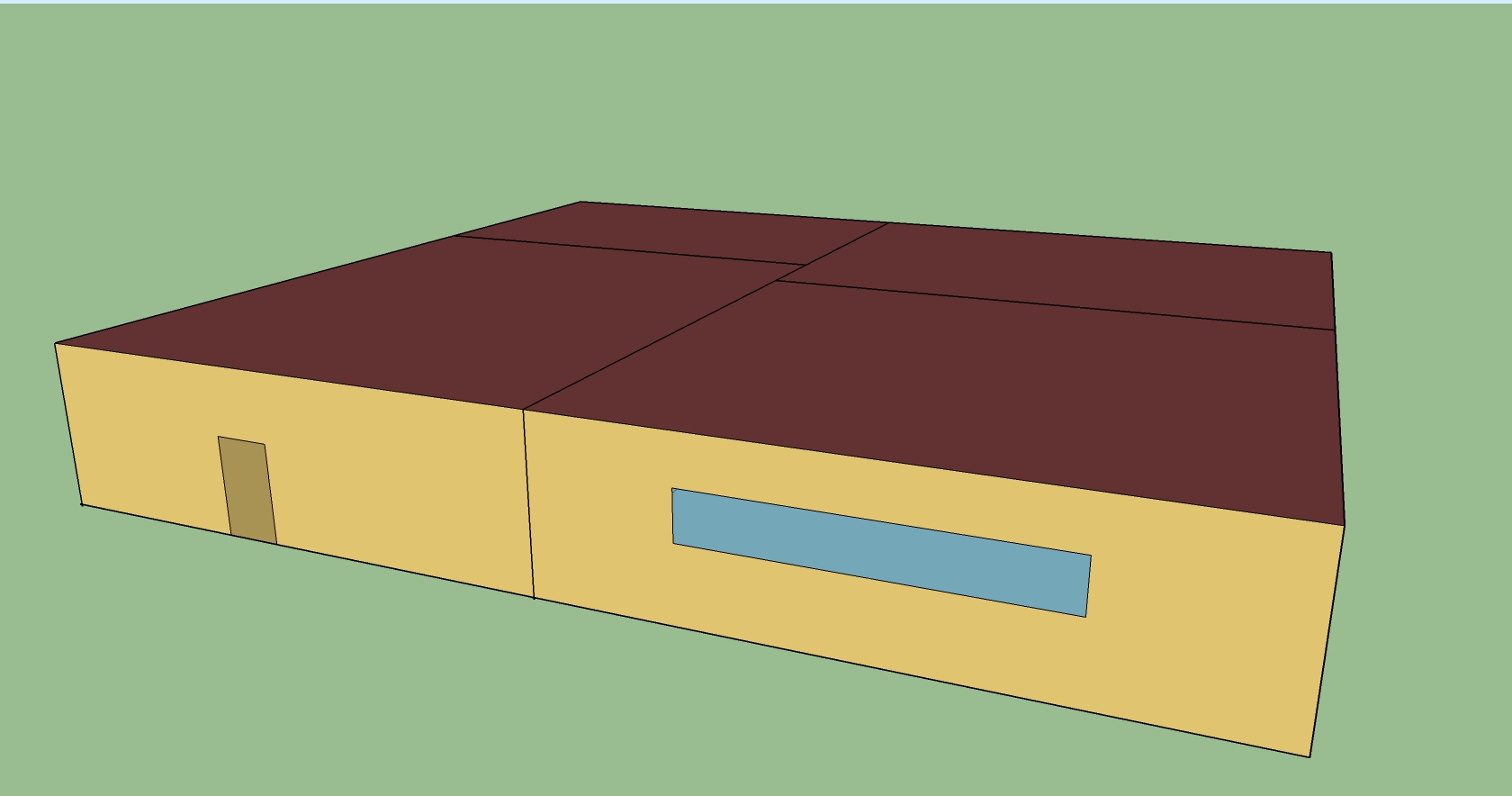}
 		\caption{Sketchup construction of the multi-zone model.}
        \label{MultizoneModel}
\end{figure}

\begin{figure}[ht!]
\centering
        \includegraphics[scale=0.4]{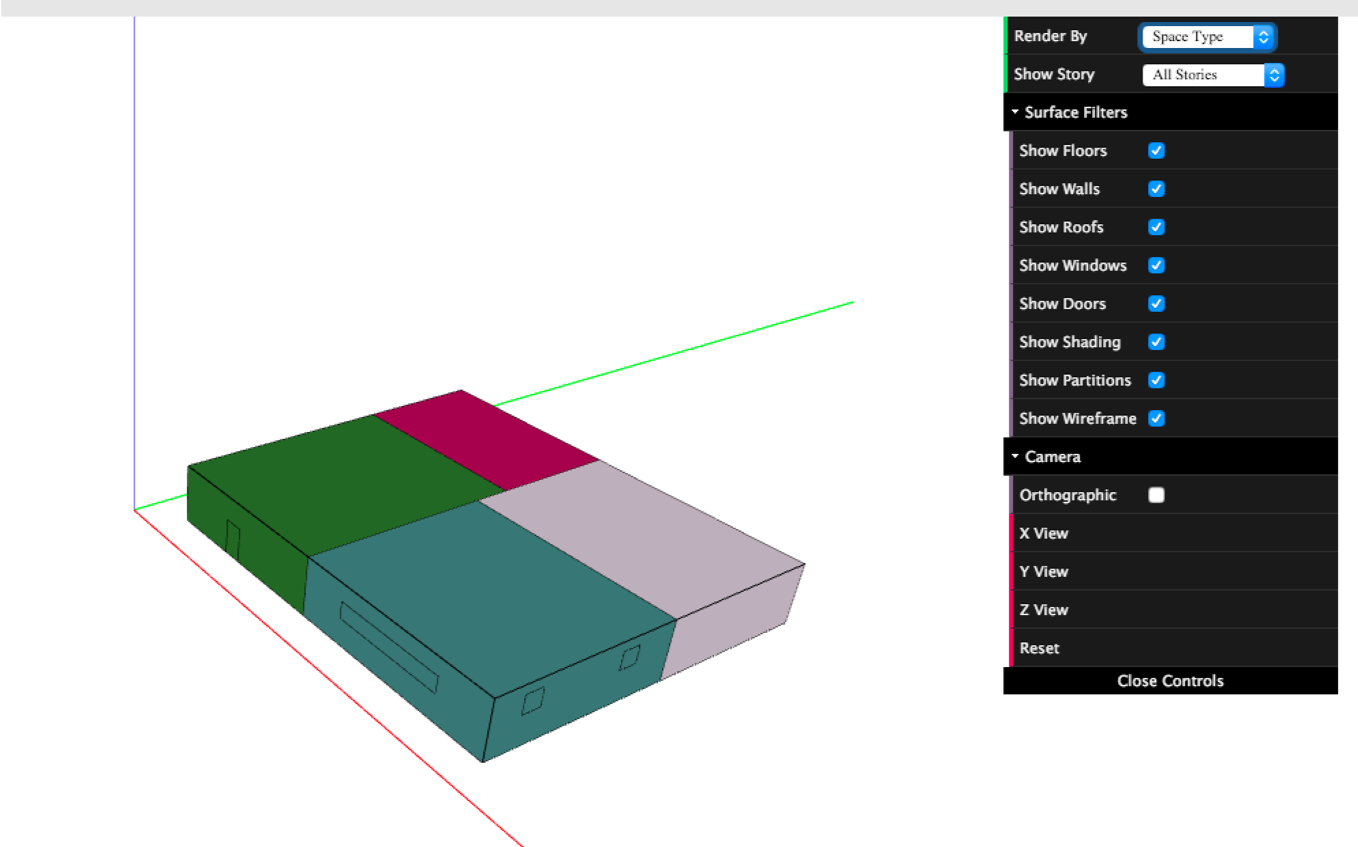}
 		\caption{Four different space types  for the multi-zone model.}
        \label{SpaceType}
\end{figure}

\begin{figure}[ht!]
\centering
        \includegraphics[scale=0.4]{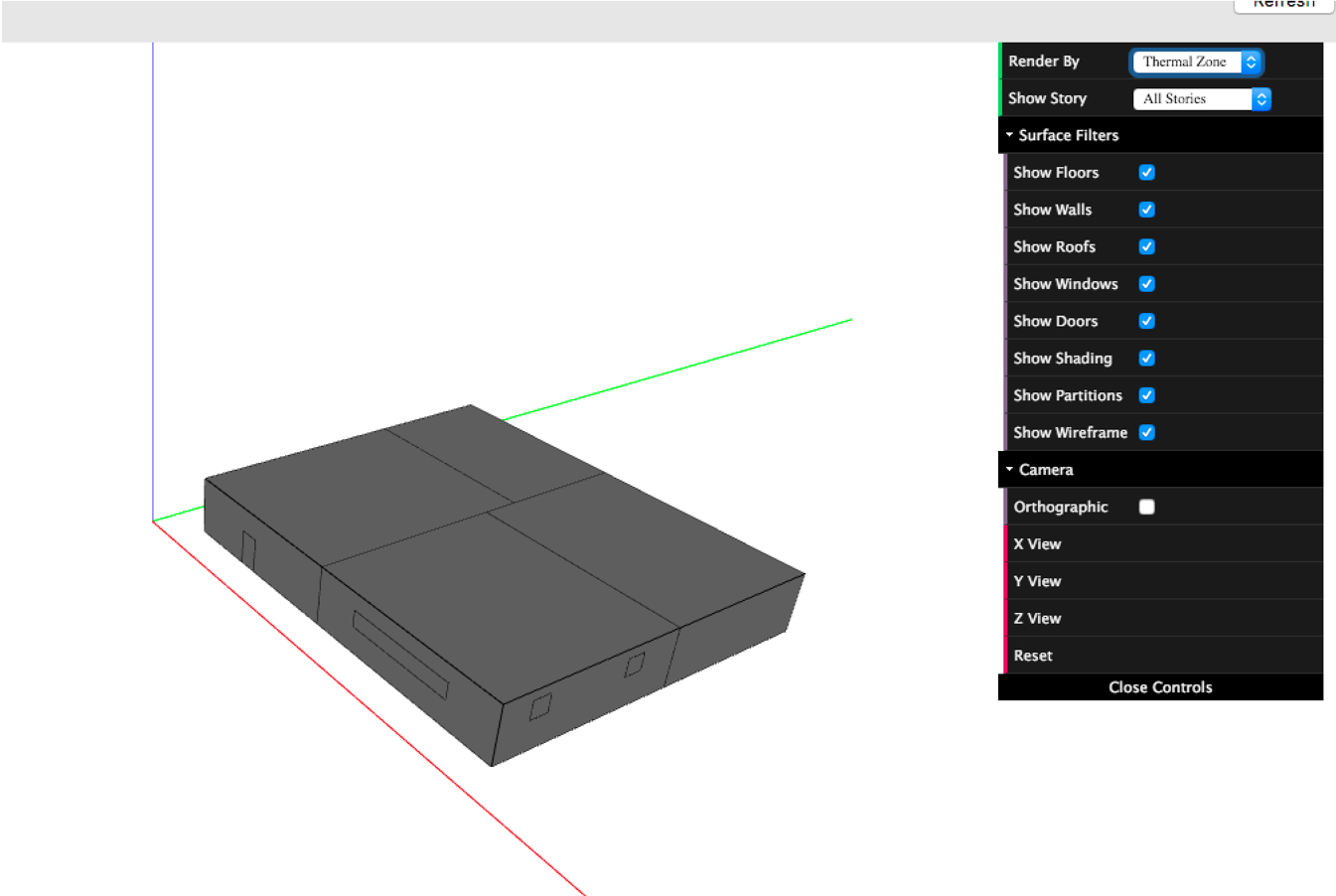}
 		\caption{Single thermal zone represengtation of a multi-zone building.}
        \label{ThermalZone}
\end{figure}

\begin{figure}[ht!]
\centering
        \includegraphics[scale=0.6]{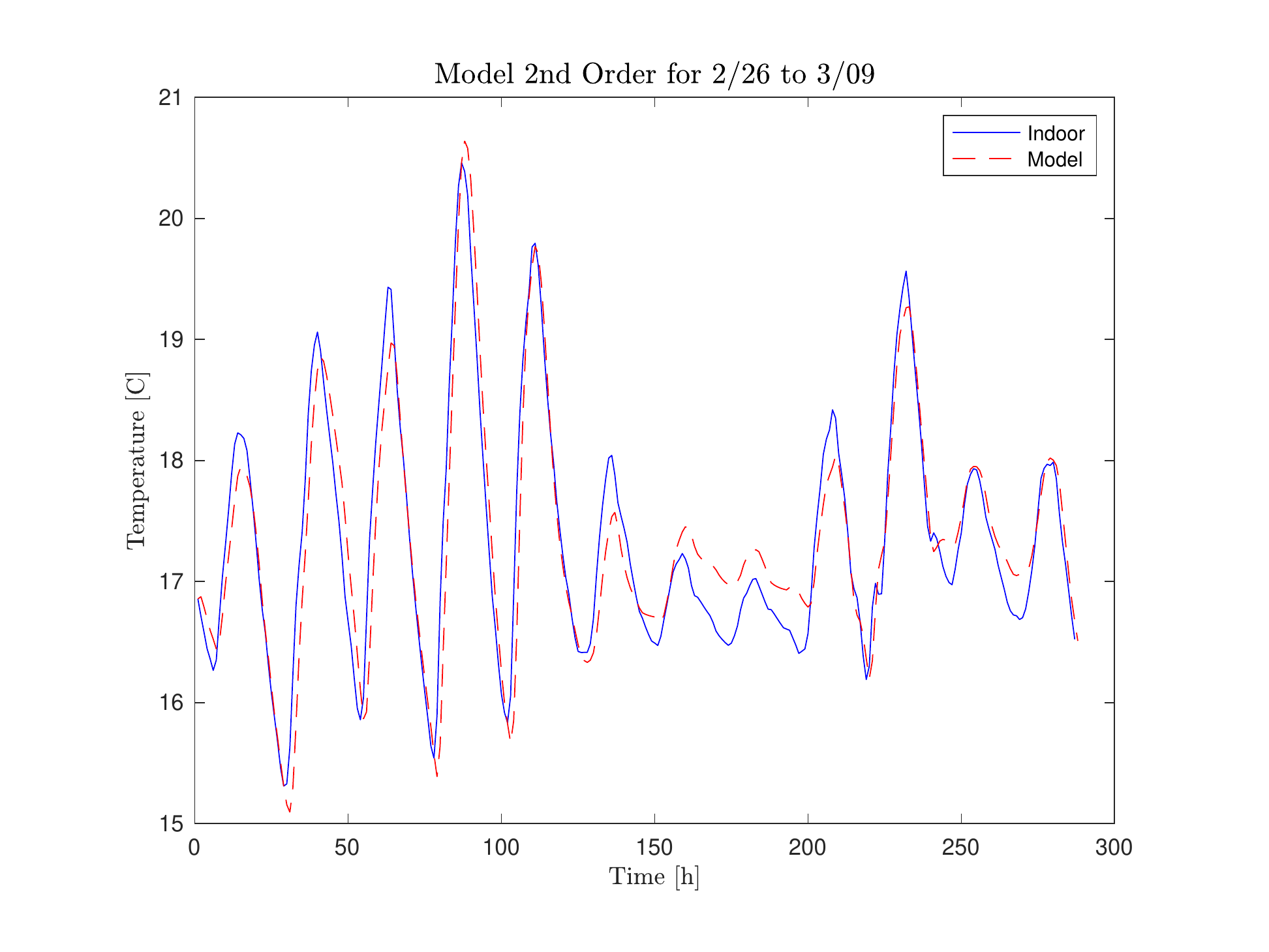}
 		\caption{Multi-zone model.}
        \label{Multizone Model}
\end{figure}

\indent In Figure \ref{Multizone Model} we  see that the model performance is similar, with a $4.1515\%$ difference from the ``true" indoor temperature obtained using EnergyPlus. Even with having one thermal zone but four different spaces in a house, we see that the model will hold. In this sense, the reduced order model recognizes the homogenized coefficients such as the thermal mass for the whole building, indicating that the modeling can be done using a systematic layered approach, where both single zone and multi zone reduced order models are constructed.\end{section}
\section{Dependence of Coefficients on Material Properties}
Building materials can affect the reduced order model coefficients substantially, illuminating their role in 
performance and efficiency of the thermal design of a building. In the appendix we present results of reduced order model coefficients with a variety of building materials, showing their effect on physical coefficients. From the results, it is evident that the variation of the thermal mass coefficient $c_1$ can be substantial, almost an order of magnitude. The damping coefficient $c_2$ and the ``thermal stiffness" $c_3$ are affected less. In all the cases, the optimal coefficients provided for a good match with the Energy Plus data.
\section{Conclusions}
\indent In this paper we proposed a reduced order modeling methodology based on Koopman operator theory. The methodology leads to linear second order zone models featuring coefficients related to global physical properties of the space, such as the thermal mass. In fact, our effort can be seen as a way to define thermal mass for a zone as the coefficient $c_1$ in the reduced order model.

\indent We tested the approach using simulated data from Energy Plus model for single and multi-zone buildings and found that optimized coefficients provide a good match of the reduced order model with the data. We also analyzed how different materials affect properties such as the thermal mass, finding that variation in the thermal mass can be very substantial depending on wall materials used.

\indent The low complexity, high accuracy reduced order models developed here can be used in development of controllers with standard actuation, but also non-standard, such as the active thermal mass actuation.


\FloatBarrier
\section*{Acknowledgments}

{We would like to thank} Jim and Beverly Zaleski for making this work possible.

%
\FloatBarrier
\section*{Funding}
This research was made possible by the generous gift from the Zaleski Foundation.

\newpage
\section{References}

\bibliographystyle{apacite}
\bibliography{references.bib}  

\begin{thebibliography}{}

\bibitem [\protect \citeauthoryear {%
Antar%
}{%
Antar%
}{%
{\protect \APACyear {2010}}%
}]{%
radiation}
\APACinsertmetastar {%
radiation}%
\begin{APACrefauthors}%
Antar, M\BPBI A.%
\end{APACrefauthors}%
\unskip\
\newblock
\APACrefYearMonthDay{2010}{}{}.
\newblock
{\BBOQ}\APACrefatitle {Thermal radiation role in conjugate heat transfer across
  a multiple-cavity building block} {Thermal radiation role in conjugate heat
  transfer across a multiple-cavity building block}.{\BBCQ}
\newblock
\APACjournalVolNumPages{Energy}{35}{8}{3508--3516}.
\PrintBackRefs{\CurrentBib}

\bibitem [\protect \citeauthoryear {%
Avci%
, Erkoc%
, Rahmani%
\BCBL {}\ \BBA {} Asfour%
}{%
Avci%
\ \protect \BOthers {.}}{%
{\protect \APACyear {2013}}%
}]{%
avci2013model}
\APACinsertmetastar {%
avci2013model}%
\begin{APACrefauthors}%
Avci, M.%
, Erkoc, M.%
, Rahmani, A.%
\BCBL {}\ \BBA {} Asfour, S.%
\end{APACrefauthors}%
\unskip\
\newblock
\APACrefYearMonthDay{2013}{}{}.
\newblock
{\BBOQ}\APACrefatitle {Model predictive HVAC load control in buildings using
  real-time electricity pricing} {Model predictive hvac load control in
  buildings using real-time electricity pricing}.{\BBCQ}
\newblock
\APACjournalVolNumPages{Energy and Buildings}{60}{}{199--209}.
\PrintBackRefs{\CurrentBib}

\bibitem [\protect \citeauthoryear {%
Boskic%
, Brown%
\BCBL {}\ \BBA {} Mezic%
}{%
Boskic%
\ \protect \BOthers {.}}{%
{\protect \APACyear {2020}}%
}]{%
Boskic2020}
\APACinsertmetastar {%
Boskic2020}%
\begin{APACrefauthors}%
Boskic, L.%
, Brown, C.%
\BCBL {}\ \BBA {} Mezic, I.%
\end{APACrefauthors}%
\unskip\
\newblock
\APACrefYearMonthDay{2020}{}{}.
\newblock
{\BBOQ}\APACrefatitle {Creating zoning approximations to building energy models
  using the Koopman operator} {Creating zoning approximations to building
  energy models using the koopman operator}.{\BBCQ}
\newblock
\APACjournalVolNumPages{Advances in Building Energy Simulation,
  Submitted}{}{}{}.
\PrintBackRefs{\CurrentBib}

\bibitem [\protect \citeauthoryear {%
Bureau%
}{%
Bureau%
}{%
{\protect \APACyear {2018}}%
}]{%
USHomes}
\APACinsertmetastar {%
USHomes}%
\begin{APACrefauthors}%
Bureau, U\BPBI S\BPBI C.%
\end{APACrefauthors}%
\unskip\
\newblock
\APACrefYearMonthDay{2018}{}{}.
\newblock
{\BBOQ}\APACrefatitle {Number of housing units in the United States from 1975
  to 2017} {Number of housing units in the united states from 1975 to
  2017}{\BBCQ}\ [\bibcomputersoftwaremanual].
\newblock
\APACrefnote{https://www.statista.com/statistics/$240267$/number-of-housing-units-in-the-united-states/}
\PrintBackRefs{\CurrentBib}

\bibitem [\protect \citeauthoryear {%
Ford%
, Pritoni%
, Sanguinetti%
\BCBL {}\ \BBA {} Karlin%
}{%
Ford%
\ \protect \BOthers {.}}{%
{\protect \APACyear {2017}}%
}]{%
ford2017}
\APACinsertmetastar {%
ford2017}%
\begin{APACrefauthors}%
Ford, R.%
, Pritoni, M.%
, Sanguinetti, A.%
\BCBL {}\ \BBA {} Karlin, B.%
\end{APACrefauthors}%
\unskip\
\newblock
\APACrefYearMonthDay{2017}{}{}.
\newblock
{\BBOQ}\APACrefatitle {Categories and functionality of smart home technology
  for energy management} {Categories and functionality of smart home technology
  for energy management}.{\BBCQ}
\newblock
\APACjournalVolNumPages{Building and Environment}{123}{}{543--554}.
\PrintBackRefs{\CurrentBib}

\bibitem [\protect \citeauthoryear {%
Georgescu%
, Eisenhower%
\BCBL {}\ \BBA {} Mezic%
}{%
Georgescu%
\ \protect \BOthers {.}}{%
{\protect \APACyear {2012}}%
}]{%
georgescu2012creating}
\APACinsertmetastar {%
georgescu2012creating}%
\begin{APACrefauthors}%
Georgescu, M.%
, Eisenhower, B.%
\BCBL {}\ \BBA {} Mezic, I.%
\end{APACrefauthors}%
\unskip\
\newblock
\APACrefYearMonthDay{2012}{}{}.
\newblock
{\BBOQ}\APACrefatitle {Creating zoning approximations to building energy models
  using the Koopman operator} {Creating zoning approximations to building
  energy models using the koopman operator}.{\BBCQ}
\newblock
\APACjournalVolNumPages{Proceedings of SimBuild}{5}{1}{40--47}.
\PrintBackRefs{\CurrentBib}

\bibitem [\protect \citeauthoryear {%
Georgescu%
\ \BBA {} Mezi{\'c}%
}{%
Georgescu%
\ \BBA {} Mezi{\'c}%
}{%
{\protect \APACyear {2015}}%
}]{%
georgescuandmezic:2015}
\APACinsertmetastar {%
georgescuandmezic:2015}%
\begin{APACrefauthors}%
Georgescu, M.%
\BCBT {}\ \BBA {} Mezi{\'c}, I.%
\end{APACrefauthors}%
\unskip\
\newblock
\APACrefYearMonthDay{2015}{}{}.
\newblock
{\BBOQ}\APACrefatitle {Building energy modeling: A systematic approach to
  zoning and model reduction using Koopman Mode Analysis} {Building energy
  modeling: A systematic approach to zoning and model reduction using koopman
  mode analysis}.{\BBCQ}
\newblock
\APACjournalVolNumPages{Energy and buildings}{86}{}{794--802}.
\PrintBackRefs{\CurrentBib}

\bibitem [\protect \citeauthoryear {%
Hazyuk%
, Ghiaus%
\BCBL {}\ \BBA {} Penhouet%
}{%
Hazyuk%
\ \protect \BOthers {.}}{%
{\protect \APACyear {2012}}%
}]{%
hazyuk2012optimal}
\APACinsertmetastar {%
hazyuk2012optimal}%
\begin{APACrefauthors}%
Hazyuk, I.%
, Ghiaus, C.%
\BCBL {}\ \BBA {} Penhouet, D.%
\end{APACrefauthors}%
\unskip\
\newblock
\APACrefYearMonthDay{2012}{}{}.
\newblock
{\BBOQ}\APACrefatitle {Optimal temperature control of intermittently heated
  buildings using Model Predictive Control: Part I--Building modeling} {Optimal
  temperature control of intermittently heated buildings using model predictive
  control: Part i--building modeling}.{\BBCQ}
\newblock
\APACjournalVolNumPages{Building and Environment}{51}{}{379--387}.
\PrintBackRefs{\CurrentBib}

\bibitem [\protect \citeauthoryear {%
Hoicka%
\ \BBA {} Parker%
}{%
Hoicka%
\ \BBA {} Parker%
}{%
{\protect \APACyear {2018}}%
}]{%
hoickaetal:2018}
\APACinsertmetastar {%
hoickaetal:2018}%
\begin{APACrefauthors}%
Hoicka, C\BPBI E.%
\BCBT {}\ \BBA {} Parker, P.%
\end{APACrefauthors}%
\unskip\
\newblock
\APACrefYearMonthDay{2018}{}{}.
\newblock
{\BBOQ}\APACrefatitle {Assessing the adoption of the house as a system approach
  to residential energy efficiency programs} {Assessing the adoption of the
  house as a system approach to residential energy efficiency programs}.{\BBCQ}
\newblock
\APACjournalVolNumPages{Energy Efficiency}{11}{2}{295--313}.
\PrintBackRefs{\CurrentBib}

\bibitem [\protect \citeauthoryear {%
Inc.%
}{%
Inc.%
}{%
{\protect \APACyear {2017}}%
}]{%
Sketchup}
\APACinsertmetastar {%
Sketchup}%
\begin{APACrefauthors}%
Inc., T.%
\end{APACrefauthors}%
\unskip\
\newblock
\APACrefYearMonthDay{2017}{}{}.
\newblock
{\BBOQ}\APACrefatitle {SketchUp Pro 2018} {Sketchup pro 2018}{\BBCQ}\
  [\bibcomputersoftwaremanual].
\newblock
\APACrefnote{https://www.sketchup.com/products/sketchup-pro}
\PrintBackRefs{\CurrentBib}

\bibitem [\protect \citeauthoryear {%
Littooy%
, Loire%
, Georgescu%
\BCBL {}\ \BBA {} Mezi{\'c}%
}{%
Littooy%
\ \protect \BOthers {.}}{%
{\protect \APACyear {2016}}%
}]{%
Littooyetal:2016}
\APACinsertmetastar {%
Littooyetal:2016}%
\begin{APACrefauthors}%
Littooy, B.%
, Loire, S.%
, Georgescu, M.%
\BCBL {}\ \BBA {} Mezi{\'c}, I.%
\end{APACrefauthors}%
\unskip\
\newblock
\APACrefYearMonthDay{2016}{}{}.
\newblock
{\BBOQ}\APACrefatitle {Pattern recognition and classification of HVAC
  rule-based faults in commercial buildings} {Pattern recognition and
  classification of hvac rule-based faults in commercial buildings}.{\BBCQ}
\newblock
\BIn{} \APACrefbtitle {Big Data (Big Data), 2016 IEEE International Conference
  on} {Big data (big data), 2016 ieee international conference on}\ (\BPGS\
  1412--1421).
\PrintBackRefs{\CurrentBib}

\bibitem [\protect \citeauthoryear {%
Masaki%
, Susuki%
, Mezic%
\BCBL {}\ \BBA {} Ishigame%
}{%
Masaki%
\ \protect \BOthers {.}}{%
{\protect \APACyear {2019}}%
}]{%
Masaki_2019}
\APACinsertmetastar {%
Masaki_2019}%
\begin{APACrefauthors}%
Masaki, I.%
, Susuki, Y.%
, Mezic, I.%
\BCBL {}\ \BBA {} Ishigame, A.%
\end{APACrefauthors}%
\unskip\
\newblock
\APACrefYearMonthDay{2019}{mar}{}.
\newblock
{\BBOQ}\APACrefatitle {An {LC}-Circuit Model for Dynamics of In-Building Heat
  Transfer across Atrium Space} {An {LC}-circuit model for dynamics of
  in-building heat transfer across atrium space}.{\BBCQ}
\newblock
\APACjournalVolNumPages{{IOP} Conference Series: Earth and Environmental
  Science}{238}{}{012012}.
\newblock
\begin{APACrefDOI} \doi{10.1088/1755-1315/238/1/012012} \end{APACrefDOI}
\PrintBackRefs{\CurrentBib}

\bibitem [\protect \citeauthoryear {%
Mauroy%
, Mezi{\'c}%
\BCBL {}\ \BBA {} Susuki%
}{%
Mauroy%
\ \protect \BOthers {.}}{%
{\protect \APACyear {2019}}%
}]{%
mauroy2019koopman}
\APACinsertmetastar {%
mauroy2019koopman}%
\begin{APACrefauthors}%
Mauroy, A.%
, Mezi{\'c}, I.%
\BCBL {}\ \BBA {} Susuki, Y.%
\end{APACrefauthors}%
\unskip\
\newblock
\APACrefYearMonthDay{2019}{}{}.
\newblock
{\BBOQ}\APACrefatitle {The Koopman Operator in Systems and Control: Theory,
  Numerics, and Applications} {The koopman operator in systems and control:
  Theory, numerics, and applications}.{\BBCQ}
\newblock
\APACjournalVolNumPages{Springer}{}{}{}.
\PrintBackRefs{\CurrentBib}

\bibitem [\protect \citeauthoryear {%
May-Ostendorp%
, Henze%
, Corbin%
, Rajagopalan%
\BCBL {}\ \BBA {} Felsmann%
}{%
May-Ostendorp%
\ \protect \BOthers {.}}{%
{\protect \APACyear {2011}}%
}]{%
may2011model}
\APACinsertmetastar {%
may2011model}%
\begin{APACrefauthors}%
May-Ostendorp, P.%
, Henze, G\BPBI P.%
, Corbin, C\BPBI D.%
, Rajagopalan, B.%
\BCBL {}\ \BBA {} Felsmann, C.%
\end{APACrefauthors}%
\unskip\
\newblock
\APACrefYearMonthDay{2011}{}{}.
\newblock
{\BBOQ}\APACrefatitle {Model-predictive control of mixed-mode buildings with
  rule extraction} {Model-predictive control of mixed-mode buildings with rule
  extraction}.{\BBCQ}
\newblock
\APACjournalVolNumPages{Building and Environment}{46}{2}{428--437}.
\PrintBackRefs{\CurrentBib}

\bibitem [\protect \citeauthoryear {%
Mezi{\'c}%
}{%
Mezi{\'c}%
}{%
{\protect \APACyear {2005}}%
}]{%
Mezic:2005}
\APACinsertmetastar {%
Mezic:2005}%
\begin{APACrefauthors}%
Mezi{\'c}, I.%
\end{APACrefauthors}%
\unskip\
\newblock
\APACrefYearMonthDay{2005}{}{}.
\newblock
{\BBOQ}\APACrefatitle {Spectral properties of dynamical systems, model
  reduction and decompositions} {Spectral properties of dynamical systems,
  model reduction and decompositions}.{\BBCQ}
\newblock
\APACjournalVolNumPages{Nonlinear Dynamics}{41}{1-3}{309--325}.
\PrintBackRefs{\CurrentBib}

\bibitem [\protect \citeauthoryear {%
Mezi{\'c}%
}{%
Mezi{\'c}%
}{%
{\protect \APACyear {2013}}%
}]{%
mezic2013analysis}
\APACinsertmetastar {%
mezic2013analysis}%
\begin{APACrefauthors}%
Mezi{\'c}, I.%
\end{APACrefauthors}%
\unskip\
\newblock
\APACrefYearMonthDay{2013}{}{}.
\newblock
{\BBOQ}\APACrefatitle {Analysis of fluid flows via spectral properties of the
  Koopman operator} {Analysis of fluid flows via spectral properties of the
  koopman operator}.{\BBCQ}
\newblock
\APACjournalVolNumPages{Annual Review of Fluid Mechanics}{45}{}{357--378}.
\PrintBackRefs{\CurrentBib}

\bibitem [\protect \citeauthoryear {%
Mezic%
}{%
Mezic%
}{%
{\protect \APACyear {2017}}%
}]{%
Mezic:2017}
\APACinsertmetastar {%
Mezic:2017}%
\begin{APACrefauthors}%
Mezic, I.%
\end{APACrefauthors}%
\unskip\
\newblock
\APACrefYearMonthDay{2017}{}{}.
\newblock
\APACrefbtitle {Dynamics of System of Systems and Applications to Net Zero
  Energy Facilities} {Dynamics of system of systems and applications to net
  zero energy facilities}\ \APACbVolEdTR{}{\BTR{}}.
\newblock
\APACaddressInstitution{University of California-Santa Barbara Santa Barbara
  United States}{University of California-Santa Barbara Santa Barbara United
  States}.
\PrintBackRefs{\CurrentBib}

\bibitem [\protect \citeauthoryear {%
Mezi{\'c}%
\ \BBA {} Banaszuk%
}{%
Mezi{\'c}%
\ \BBA {} Banaszuk%
}{%
{\protect \APACyear {2004}}%
}]{%
mezic2004comparison}
\APACinsertmetastar {%
mezic2004comparison}%
\begin{APACrefauthors}%
Mezi{\'c}, I.%
\BCBT {}\ \BBA {} Banaszuk, A.%
\end{APACrefauthors}%
\unskip\
\newblock
\APACrefYearMonthDay{2004}{}{}.
\newblock
{\BBOQ}\APACrefatitle {Comparison of systems with complex behavior} {Comparison
  of systems with complex behavior}.{\BBCQ}
\newblock
\APACjournalVolNumPages{Physica D: Nonlinear Phenomena}{197}{1-2}{101--133}.
\PrintBackRefs{\CurrentBib}

\bibitem [\protect \citeauthoryear {%
of~the U.S. Department~of Energy%
}{%
of~the U.S. Department~of Energy%
}{%
{\protect \APACyear {2017}}%
}]{%
Openstudio}
\APACinsertmetastar {%
Openstudio}%
\begin{APACrefauthors}%
of~the U.S. Department~of Energy, N\BPBI L.%
\end{APACrefauthors}%
\unskip\
\newblock
\APACrefYearMonthDay{2017}{}{}.
\newblock
{\BBOQ}\APACrefatitle {OpenStudio} {Openstudio}{\BBCQ}\
  [\bibcomputersoftwaremanual].
\newblock
\APACrefnote{https://www.openstudio.net/}
\PrintBackRefs{\CurrentBib}

\bibitem [\protect \citeauthoryear {%
Susuki%
, Mezic%
, Raak%
\BCBL {}\ \BBA {} Hikihara%
}{%
Susuki%
\ \protect \BOthers {.}}{%
{\protect \APACyear {2016}}%
}]{%
susuki2016applied}
\APACinsertmetastar {%
susuki2016applied}%
\begin{APACrefauthors}%
Susuki, Y.%
, Mezic, I.%
, Raak, F.%
\BCBL {}\ \BBA {} Hikihara, T.%
\end{APACrefauthors}%
\unskip\
\newblock
\APACrefYearMonthDay{2016}{}{}.
\newblock
{\BBOQ}\APACrefatitle {Applied Koopman operator theory for power systems
  technology} {Applied koopman operator theory for power systems
  technology}.{\BBCQ}
\newblock
\APACjournalVolNumPages{Nonlinear Theory and Its Applications,
  IEICE}{7}{4}{430--459}.
\PrintBackRefs{\CurrentBib}

\bibitem [\protect \citeauthoryear {%
\APACciteatitle {What’s the Difference Between Insulation and Thermal
  Mass?}}{%
\APACciteatitle {What’s the Difference Between Insulation and Thermal
  Mass?}}{%
{\protect \APACyear {2018}}%
}]{%
InsulationvsThermal}
\APACinsertmetastar {%
InsulationvsThermal}%
{\BBOQ}\APACrefatitle {What’s the Difference Between Insulation and Thermal
  Mass?} {What’s the difference between insulation and thermal mass?}{\BBCQ}\
  [\bibcomputersoftwaremanual].
\newblock
\APACrefYearMonthDay{2018}{}{}.
\newblock
\begin{APACrefURL}
  \url{http://letthesunwork.com/shelter/insulationthermalmass.htm}
  \end{APACrefURL}
\PrintBackRefs{\CurrentBib}

\end{thebibliography}

\newpage
\begin{section}{Appendix: Variation of Coefficients of the Reduced Order Model with Different Materials}
\begin{itemize}
\item Standard Model
\end{itemize}

\indent  We created  test models in OpenStudio with  different structural materials (steel and brick) and different  wall material.  The size of the test building is  $11.86m\times 13.99m\times 4.57m$ and will have three windows and one door. 
\begin{figure}[ht!]
\centering
        \includegraphics[scale=0.2]{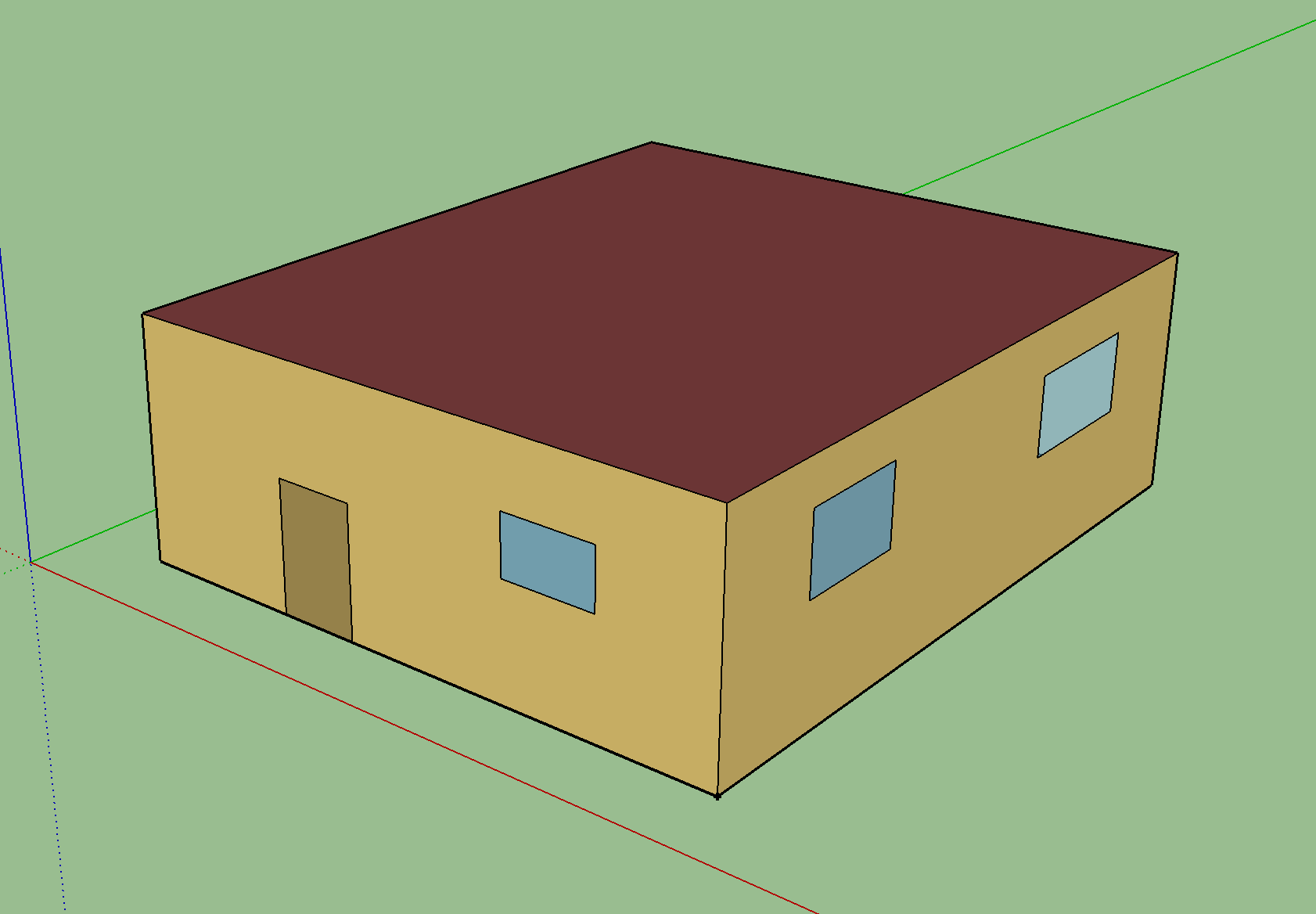}
 		\caption{The  model used in this section.}
        \label{StadardModel}
\end{figure}

\indent We used the total of 288 hourly data points.  
In the tables below we show Energy Plus model construction settings.
\begin{table}[h!]
\centering
\begin{tabular}{|c|c|c|}
\hline
Name & Material & External Wall Setting \\ \hline
Standard Model & \begin{tabular}[c]{@{}c@{}}1/2in gypsum\\ Wall insulation {[}39{]}\\ MAT-sheath\end{tabular} & Steel-framed \\ \hline
\end{tabular}
\caption{Standard model construction settings.}
\label{Table:BaseModelSetup}
\end{table}\\

For  the standard model, we observed: $c1=0.4350$, $c2=10.2650$, $c3=2.2750$. With an error of $6.1112\%$ and the model read
\begin{equation} \label{StandardModelMatrix} 
\dot{x}= \begin{bmatrix} 0 & 1 \\ -5.2299 & -23.5977 \end{bmatrix}x + 
\begin{bmatrix}
0 \\ 2.2989
\end{bmatrix}u + 
\begin{bmatrix}
1 \\ 0
\end{bmatrix}
\frac{1.10}{0.4350}
\end{equation}
\pagebreak 
\begin{itemize}
\item Brick Model
\end{itemize}\

\indent The brick design used  four different cases of materials showm in table \ref{Brick Material Table}  
\begin{table}[]
\centering
\begin{tabular}{|c|c|c|}
\hline
Name & Material & External Wall Setting \\ \hline
Brick Model & \begin{tabular}[c]{@{}c@{}}1in stucco\\ 8in concrete\\ 1/2 gypsum\\ Wall insulation {[}40{]}\end{tabular} & Brick-framed \\ \hline
Brick and Insulation & \begin{tabular}[c]{@{}c@{}}1in stucco\\ 8in concrete\\ 1/2 gypsum\\ Wall insulation {[}40{]}\\ \textbf{Wall insulation{[}40{]}}\end{tabular} & Brick-framed \\ \hline
Brick, Insulation, and Gypsum & \begin{tabular}[c]{@{}c@{}}1in stucco\\ 8in concrete\\ 1/2 gypsum\\ Wall insulation {[}40{]}\\ \textbf{Wall insulation{[}40{]}}\\ \textbf{1/2in gypsum}\end{tabular} & Brick-framed \\ \hline
Brick, Insulation, and Concrete & \begin{tabular}[c]{@{}c@{}}1in stucco\\ 8in concrete\\ 1/2 gypsum\\ Wall insulation {[}40{]}\\ \textbf{Wall insulation{[}40{]}}\\ \textbf{8in concrete}\end{tabular} & Brick-framed \\ \hline
\end{tabular}
\caption{Brick case study construction layout.}
\label{Brick Material Table}
\end{table}

\begin{figure} [h!]
    \centering
        \includegraphics[width=0.75\textwidth]{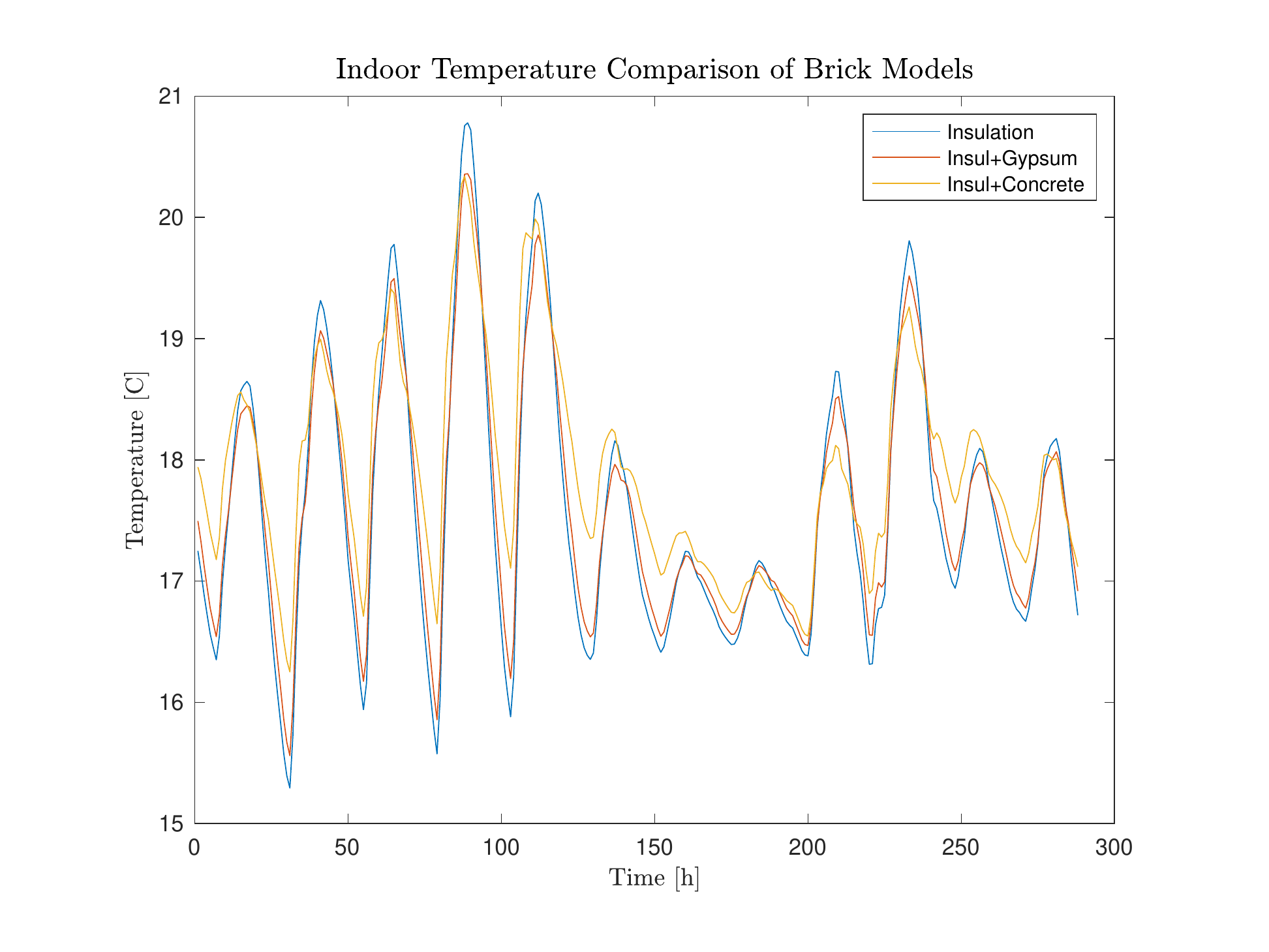}
 		\caption{Comparison of the indoor temperature for the three different internal structures of the model.}
        \label{IndoorComparethree}
\end{figure}
\begin{figure} [h]
    \centering
        \includegraphics[width=0.75\textwidth, angle = 270]{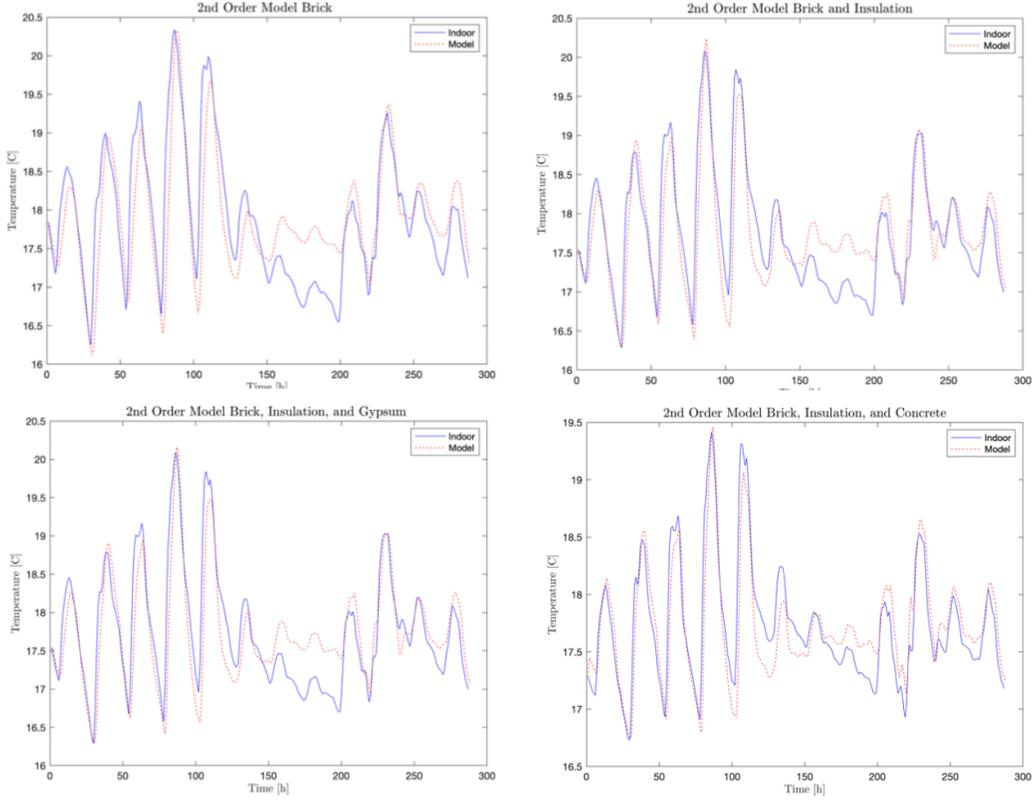}
 		\caption{ROM plot of the four test cases with a brick construction.}
        \label{BrickCompareDifferentTypes}
\end{figure}

For the pure  brick model the coefficients  were found to be, $c1=0.3800$,$c2=15.2800$, $c3=3.2550$ and $c4=1.1$. Error=7.7222\%, and the model read
\begin{equation} \label{brickMatrixAll} 
\dot{x}= \begin{bmatrix} 0 & 1 \\ -8.5658 & -40.2105 \end{bmatrix}x + 
\begin{bmatrix}
0 \\ 2.6316
\end{bmatrix}u +
\begin{bmatrix}
1 \\ 0
\end{bmatrix}
 \frac{1.10}{0.3800}
\end{equation}\

For the brick mode with added insulation the coefficients were found to be, $c1=0.1240$,     $c2=7.7750$, $c3=4.6750$ and $c4=1.1$. Error=6.0809\% and the model reads
\begin{equation} \label{brickMatrixINSUL} 
\dot{x}= \begin{bmatrix} 0 & 1 \\ -37.7016& -62.7016 \end{bmatrix}x + 
\begin{bmatrix}
0 \\ 8.0645
\end{bmatrix}u +
\begin{bmatrix}
1 \\ 0
\end{bmatrix}
 \frac{1.10}{0.1240}
\end{equation}\
For the brick mode with added insulation and gypsum teh coefficients  were found to be, $c1=0.1400$, $c2=8.950$, $c3=4.7650$ and $c4=1.1$. Error=5.9672\% and the model read
\begin{equation} \label{brickMatrixINSUL+GYPS} 
\dot{x}= \begin{bmatrix} 0 & 1 \\ -34.0307 & -64.1786 \end{bmatrix}x + 
\begin{bmatrix}
0 \\ 8.33
\end{bmatrix}u +
\begin{bmatrix}
1 \\ 0
\end{bmatrix}
 \frac{1.10}{0.1400}
\end{equation}\
For the brick mode with added insulation and concrete the coefficients  were found to be, $c1=0.0600$, $c2=6.3350$, $c3=7.3400$ and $c4=1.1$. Error=3.3882\% and the model reads
\begin{equation} \label{brickMatrixINSUL+CONCRETE} 
\dot{x}= \begin{bmatrix} 0 & 1 \\ -122.3330 & -105.5833 \end{bmatrix}x + 
\begin{bmatrix}
0 \\ 16.667
\end{bmatrix}u +
\begin{bmatrix}
1 \\ 0
\end{bmatrix}
 \frac{1.10}{0.0600}
\end{equation}\

\end{section}

\end{document}